\documentclass[aps,prx,twocolumn,showpacs,superscriptaddress,groupedaddress,floatfix,longbibliography]{revtex4-2}
\usepackage{color}
\usepackage{mathtools}
\usepackage{latexsym}
\usepackage{graphicx}
\usepackage{float}
\usepackage{dcolumn}
\usepackage{bm}
\usepackage{amssymb}
\usepackage{amsmath}
\usepackage{mathtools}
\usepackage[space]{grffile}
\usepackage{xcolor}

\usepackage[colorinlistoftodos]{todonotes}

\begin{document}



\title{Long-lived period-doubled edge modes of interacting and disorder-free Floquet spin chains}
\author{Daniel J. Yates$^{1}$} \author{Alexander G. Abanov$^{2,3}$}
\author{Aditi Mitra$^{1}$} \affiliation{$^{1}$Center for Quantum
  Phenomena, Department of Physics, New York University, 726 Broadway,
  New York, NY, 10003, USA\\ $^{2}$Simons Center for Geometry and
  Physics, Stony Brook, NY 11794, USA\\ $^{3}$Department of Physics
  and Astronomy, Stony Brook University, Stony Brook, NY 11794, USA}
\date{\today}

\maketitle

\section*{Abstract}
 Floquet spin chains have been a venue for understanding topological states of matter that are qualitatively
  different from their static counterparts by, for example, hosting  $\pi$ edge modes that show stable period-doubled dynamics.
  However the stability of these edge modes to interactions has traditionally required the system to be many-body localized
  in order to suppress heating. In contrast, here we show that even in the absence of disorder, and in the presence
  of bulk heating, $\pi$ edge modes are long lived. Their lifetime is extracted from exact diagonalization and is found to
  be non-perturbative in the interaction strength. A tunneling estimate for the lifetime is obtained by mapping the stroboscopic
  time-evolution to dynamics of a single particle in Krylov subspace. In this subspace, the $\pi$ edge mode manifests as the
  quasi-stable edge mode of an inhomogeneous Su-Schrieffer-Heeger model whose dimerization vanishes in the bulk of the Krylov chain.
  
\section*{Introduction}
Periodically driven or Floquet systems are promising venues for identifying states of matter that
do not exist in thermal equilibrium \cite{OkaRev}.
Perhaps the most striking among these athermal states are Floquet systems with new topological properties and bulk boundary
correspondences that require defining new topological invariants \cite{RudnerRev,Sondhi20}.
For example, while a static two-band system may be characterized by an integer
$\mathbb{Z}$ valued or a $\mathbb{Z}_2$ valued topological invariant \cite{Ryu10,QiRev11,Bernevigbook}, two-band free fermion Floquet
systems require at least one more topological
invariant so that the system has a $\mathbb{Z}\times \mathbb{Z}$ or a
$\mathbb{Z}_2 \times \mathbb{Z}_2$ classification \cite{Carpentier15,Roy16,Roy17a}. The additional topological invariant arises due to
the periodic nature of the Floquet spectrum. In particular, since the
energy in a periodically driven system is conserved only up to integer multiples of
the drive frequency $\Omega$, one has quasi-energy bands rather than energy bands. These
quasi-energy bands span a Floquet Brillouin zone (FBZ),
$\epsilon \in \left[-\Omega/2, \Omega/2\right]$, with the boundaries of the FBZ
being continuous. The additional topological invariant characterizes
edge modes that can reside at the
zone boundaries $\epsilon=\pm \Omega/2$.

For two dimensional systems, the additional topological invariant leads to the anomalous Floquet
insulator where a bulk has zero Chern number, yet chiral edge modes propagate at the boundary \cite{Kitagawa10,Rudner13,Nathan17,Titum20}.
For one dimensional (1d) systems, the edge modes that reside at the Floquet zone boundary
are known as $\pi$ edge modes \cite{Zoller11,Sen13,Delplace14}. Since these modes have a periodicity which is twice that of the drive period,
they are examples of boundary time crystals \cite{Sacha2018,Else2019,Khemani2019}. Of course, the $\mathbb{Z}\times \mathbb{Z}$ or a
$\mathbb{Z}_2 \times \mathbb{Z}_2$ classification implies that $0$ and $\pi$ edge modes
can exist separately, or together, leading to a rich phase diagram \cite{Khemani16,Kyser16,Kyser-I,Kyser-II}.

The $\pi$ edge modes occurring in most 1d systems \cite{Zoller11,Brandes12,Sen13,Delplace14}
also have their origin in what are known as strong $\pi$ modes (SPM) \cite{Yates19}, whose precise
definition we now give.
Denoting the $\pi$ strong mode as $\Psi_{\pi}$, these
are operators that have the property that they anti-commute with a discrete, say $\mathbb{Z}_2$ symmetry, which
we denote by $\mathcal {D}$. Thus,
$\{\Psi_{\pi}, \mathcal{D}\}=0$. In addition, the SPM anti-commutes with the Floquet unitary $U$
that generates stroboscopic time-evolution, $\{\Psi_{\pi},U\} \approx 0$. The symbol $\approx$ is to represent the
fact that the anti-commutation is strictly speaking obeyed only in the thermodynamic limit. For a finite size system of
length $L$, $\{\Psi_{\pi},U\} \propto |u|^L$, where $|u|<1$, so that the anti-commutator is suppressed exponentially in
the system size. The third feature of the SPM is that it is a local operator with the property $\Psi_\pi^2=O(1)$.
The existence of a SPM
implies an eigenspectrum phase
where each eigenstate $|n\rangle$ of a certain parity has a pair $\Psi_\pi |n\rangle$ of the opposite parity, with the quasi-energies of the
two states separated by $\pi/T$, where $T=2\pi/\Omega$ is the period of the drive. Note that the above definition of SPMs
can be generalized to more complex $2\pi/k$ edge modes where $k$ is an integer \cite{Sreejith16}.

When interactions are included, these operators no longer exactly anti-commute with $U$ in the thermodynamic
limit, and therefore acquire a lifetime. Yet a fascinating aspect of these edge modes, which is
the central topic of this paper, is that the lifetime far exceeds bulk heating times \cite{Yates19}.
These quasi-stable edge modes that almost, rather than exactly anti-commute
with $U$, are referred to as almost strong $\pi$ modes (ASPM). When disorder is present, the expectation is that many-body localization
will make the SPM stable to interactions at all times \cite{Huse14,Bahri15,Khemani16,Else16a,Potirnich17,Potter18}.
This paper, in contrast, concerns the stability of $\pi$ modes for disorder-free chains, and determines
how their lifetime depends on interactions. There are of course other contexts, besides ASPMs, where disorder-free Floquet systems can be
stable to interactions for long times \cite{Sheng17,Haldar18,Haldar21,Chandran16,Natsheh21a,Natsheh21b}.

An analogous definition exists for a strong zero mode (SZM) $\Psi_0$ \cite{Kitaev01,Fendley12,Fendley14,Fendley16}.
This is a local operator that obeys all the
above properties except that $\Psi_0$ commutes with the Floquet unitary in the thermodynamic limit $\left[\Psi_0, U\right]\approx 0$.
Thus existence of a SZM implies an eigenspectrum phase where the entire spectrum of $U$ is at least doubly degenerate. Adding
interactions makes the SZM into a quasi-stable almost strong zero mode (ASZM).
Since SZMs and ASZMs have by now been discussed in detail in static systems \cite{Nayak17,Fendley17,Parker19,Laumann20,Yates20,Yates20a},
in this paper we will only focus on SPMs and ASPMs.
A central goal of this paper is to establish how the lifetime of the
  $\pi$ mode depends on interactions. In the process we present Krylov time-evolution as a tool
  for studying Floquet dynamics. This approach
 has so far only been employed for static Hamiltonians \cite{Recbook,Gorsky19,Avdoshkin19,Yates20,Yates20a}.
 Here we show how Krylov techniques can be used to extract tunneling times of quasi-stable modes of driven systems.

We will study the stroboscopic dynamics of an open spin-$1/2$ chain of length $L$. The
dimension of the Hilbert space of the problem is $2^{L}$.
The natural numerical method of choice is exact diagonalization (ED), and we are limited to
a system size of $L=14$. Since we are
interested in extracting lifetimes in the thermodynamic limit of
$L \rightarrow \infty$, ED can be restrictive, and alternate numerical and analytical methods
are needed. Thus the results presented here, besides being an explication of the unusual physics of ASPMs,
also provide a roadmap for developing methods for understanding slow, system size independent dynamics.

\section*{Results and Discussion}

{\bf Model:} We will study an open chain of length $L$ whose
stroboscopic time-evolution is generated by the following Floquet unitary
\begin{equation}
  U = e^{-i\frac{T}{2} J_z H_{zz}} e^{-i \frac{T}{2} J_x H_{xx}}e^{-i \frac{T}{2} g H_z}, \label{eqU1}
\end{equation}
where
\begin{align}
  H_z = \sum_{i=1}^L \sigma^z_i;\, H_{xx} = \sum_{i=1}^{L-1}\sigma^x_i \sigma^x_{i+1};\,  H_{zz} = \sum_{i=1}^{L-1} \sigma^z_i \sigma^z_{i+1}.
\end{align}
Above $\sigma^{x,y,z}$ are the Pauli matrices, and
in what follows we set $J_x=1$. Eq.~\eqref{eqU1} is an example of
a ternary drive where during one period, a magnetic field of
strength $T g$ is applied, followed by the application of a nearest neighbor exchange
interaction of strength $T$ along the $x$ direction, and this is followed by the
application of a nearest neighbor exchange interaction of strength $T J_z$ along the $z$-direction.
The Floquet unitary has a discrete symmetry as it commutes with
$\mathcal{D} = \sigma_1^z \dots \sigma_L^z$.

When
$J_z=0$, the model can be mapped to free fermions through a Jordan-Wigner transformation.
In this free limit any operator can be expanded in
a basis of $2L$ Majorana fermions or Pauli string operators. Thus when $J_z=0$, the dimension of the space needed to
diagonalize the problem is only $2L$ rather than the dimension of the full Hilbert space $2^L$.
The existence of this free limit
has been valuable for identifying strong modes. In fact,
at the special point of $T g =\pi$, the Floquet unitary for $J_z=0$ reduces to
$U = (-i)^L{\mathcal D}  e^{-i \frac{T}{2} J_x H_{xx}}$  \cite{Khemani16}. At this special point,
the SPM is simply the Pauli spin operator on the first site
$\Psi_\pi = \sigma^x_1$. It is straightforward to check that $\sigma^x_1$ has
all the properties of a SPM as summarized in the Introduction. It is a local operator with $\Psi_{\pi}^2=1$,
it anti-commutes with ${\mathcal D}$, and it anti-commutes with $U$.
Even away from this special point,
although the SPM is a more complex operator, it continues to
have the property that it has a non-zero overlap with $\sigma^x_1$ \cite{Sen13,Yates19}.

Including interactions, i.e, $J_z\neq 0$, as anticipated in the Introduction, the SPM changes to an ASPM. Moreover, this operator
continues to have an overlap with $\sigma^x_1$ for weak interactions. Therefore, an efficient way to determine whether
the system hosts these special edge modes is through the study of the following autocorrelation
function
\begin{subequations}\label{eqAinf}
\begin{align}
  A_{\infty}(n T) &= \frac{1}{2^L}{\rm Tr} \biggl[\sigma^x_1(n T) \sigma^x_1(0)\biggr],\\
  \sigma^x(n T) &= \left[U^{\dagger}\right]^n \sigma^x_1 U^n .
\end{align}
\end{subequations}
We only study the dynamics at stroboscopic times, so $n$ is an integer.
Moreover, $A_{\infty}$ is an infinite temperature average as the trace is over the entire Hilbert space.
Thus this quantity is a highly out of equilibrium measure of the system dynamics. When an ASPM or an ASZM is non-existent
or if we replace $\sigma^x_1$ by an operator deep in the bulk of the chain, $A_{\infty}$ will decay to zero within a few
drive cycles.
Throughout this paper, all numerical results will be for $g=0.3$ and $T=8.25$, but for different values of $J_z$.
For this value of $g$, $T$, when $J_z=0$ we have a SPM. When $J_z$ is non-zero, the SPM changes to an ASPM.

\begin{figure}[ht]
\includegraphics[width = 0.49\textwidth]{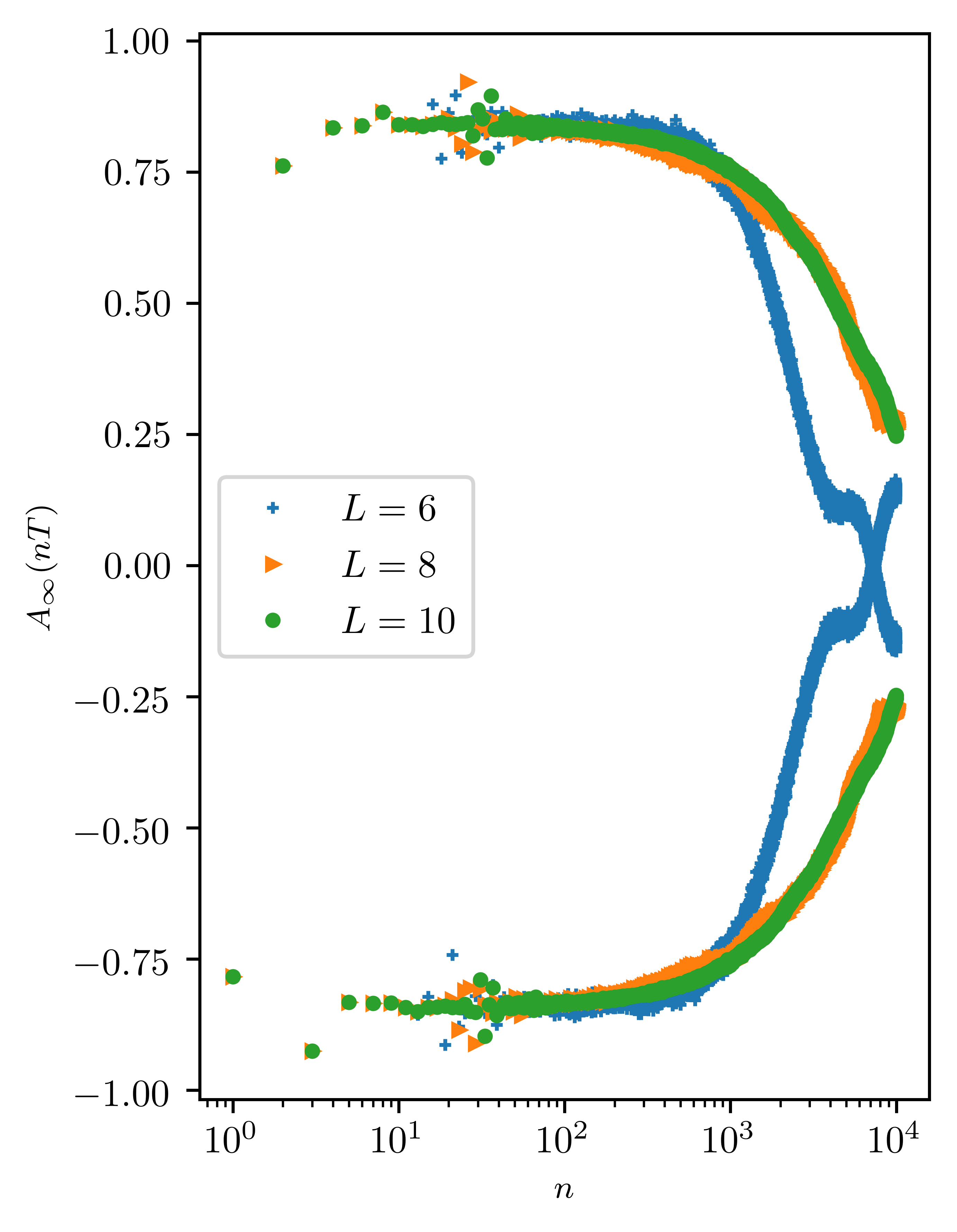}
\caption{\label{fig1a} {\bf System size dependence of the autocorrelation function}. The autocorrelation function $A_{\infty}$ at stroboscopic times
  $n T$, where $T$ is the drive period, and $n$ is an integer. The results are for a
  magnetic field of strength $g=0.3$, interaction of strength $J_z=0.01$, and period $T=8.25$,
  for three different system sizes $L= 6,8,10$. The $x$-axis is on a logarithmic scale.
  The system hosts an almost strong $\pi$ mode (ASPM) as is clearly visible by the flipping of the sign of $A_{\infty}$ between neighboring
  stroboscopic times. For these parameters the ASPM
  decays at the same time for $L=8$ and $L=10$. Thus the thermodynamic limit for the lifetime is already reached for
  $L=8$, with this lifetime being approximately $10^4$ drive cycles.
}
\end{figure}

{\bf Exact Diagonalization (ED):} We first present results for the stroboscopic time-evolution of $A_{\infty}$ from ED.
Fig.~\ref{fig1a} plots $A_{\infty}$ for $J_z=0.01$, for three different system sizes
$L=6,8,10$. The $x$-axis which denotes the stroboscopic times is on a logarithmic
scale. One finds that $A_{\infty}$ flips sign between neighboring stroboscopic times, thus we have an ASPM. Moreover,
for small system sizes, as $L$ increases, the lifetime of the ASPM increases (compare $L=6$ with $L=8$ in the figure), but eventually
the lifetime reaches a system size independent value. For the chosen parameters, the system size independent lifetime is
reached by $L=8$ as the plots for $L=8$ and $L=10$ nearly coincide. We refer to this system size independent lifetime as the lifetime in the
thermodynamic limit.
Moreover, for $J_z=0.01$, this lifetime is around $10^4$ drive cycles. Note that
the high frequency limit requires $T \ll 1$, and since we have $T=8.25$, we are far from the high frequency limit.
Thus, the bulk is in fact heating, as expected for a periodically driven interacting, and disorder-free
system \cite{Hyungwon14,Lazarides14,DAlessio14,Ponte15,Bukov16}.
The evidence from ED for bulk heating was already presented in Ref.~\onlinecite{Yates19}, where it
was shown that the autocorrelation function for bulk operators decays to zero within a few drive cycles, and that
the entropy density rapidly approaches the maximum possible value (accounting for finite size effects). Further below we
 discuss the signature of bulk heating in Krylov subspace. We also extend the results of Ref.~\onlinecite{Yates19}
 by extracting the interaction dependence of the
 lifetime of the $\pi$-mode using several different approaches: ED, Krylov dynamics, and domain wall counting.

\begin{figure}[ht]
\includegraphics[width = 0.49\textwidth]{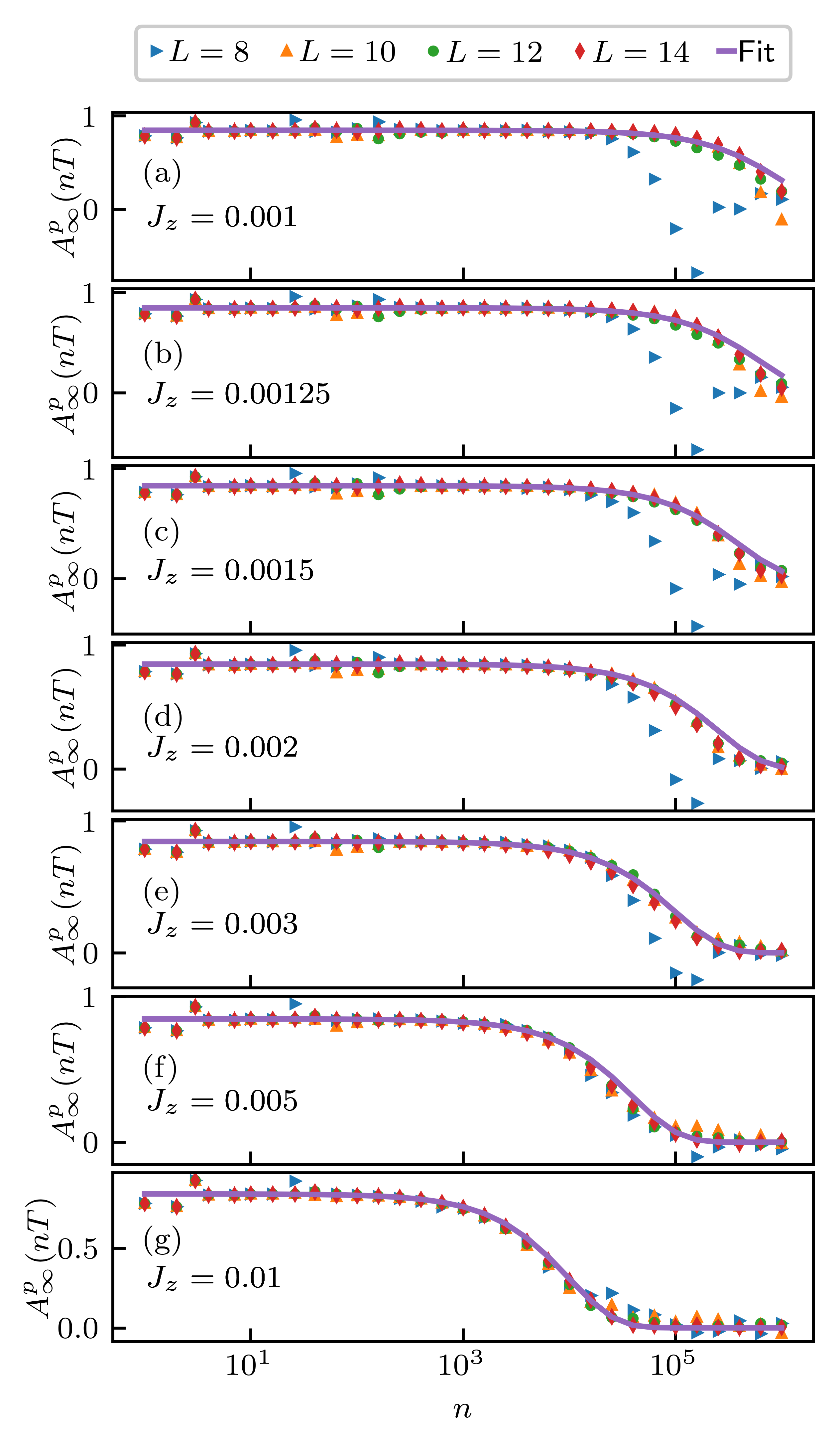}
\caption{\label{fig2a} {\bf Interaction and system size dependence of the autocorrelation function}.
  The autocorrelation function accounting for the sign-change between neighboring strobosocopic times
  $A_{\infty}^p(nT) =(-1)^n A_{\rm \infty}(n T)$ is plotted for stroboscopic and logarithmically separated times $n T$, where $T$ is the period
  and $n$ is an integer. The results are for a magnetic field of
  strength $g=0.3$, period $T=8.25$, and for
  different interaction strengths $J_z$. The $x$-axis is on a logarithmic scale. From panels (a-g) $J_z$ increases. Panel (g)
  corresponds to $J_z=0.01$ and the corresponding time evolution for $A_{\infty}(n T)$ for linearly separated times is shown in Fig.~\ref{fig1a}.
 Also shown is a fit (purple line) to an exponential $A_{\infty}^p(n T)=e^{-\Gamma n T}A_{\infty}^p(10 T)$, where $\Gamma$ is a decay rate.
}
\end{figure}

We now discuss how the lifetime in the thermodynamic limit depends on $J_z$.
Fig.~\ref{fig2a} shows the autocorrelation function accounting for its period-doubled behavior
$A^p_{\infty}(nT)=(-1)^nA_{\infty}(nT)$. Since the sign-flipping of $A_{\infty}$ is absorbed by the $(-1)^n$ factor, $A^p_{\infty}$ has a smoother
behavior in time. We plot $A^p_{\infty}$ for different $J_z$ and for system sizes $L=8,10,12,14$.
The smallest possible $J_z$ we can study is $J_z=0.001$ because for $J_z$ values
smaller than this, the system size needed for the lifetime to become $L$ independent, is larger than $L=14$. While the stroboscopic times
in Fig.~\ref{fig1a} were linearly separated, the stroboscopic times in Fig.~\ref{fig2a} are logarithmically separated as the
lifetimes increase dramatically with decreasing $J_z$, and linearly separated points are numerically too costly to compute.

Fig.~\ref{fig2a} clearly shows that as $J_z$ increases (panels (a-g)), the lifetime of the ASPM decreases. In fact the thermodynamic lifetimes
are already reached for $L=8$ when $J_z=0.01$ as the plots for all the
four system sizes lie on top of each other (panel (g)).
Recall that Fig.~\ref{fig1a} is a more detailed version of $A_{\infty}$ for precisely this value of $J_z$,
where the stroboscopic times are linearly separated, and one smaller system size, $L=6$, is shown in order to highlight the system
size dependence.

\begin{figure}[ht]
\includegraphics[width = 0.49\textwidth]{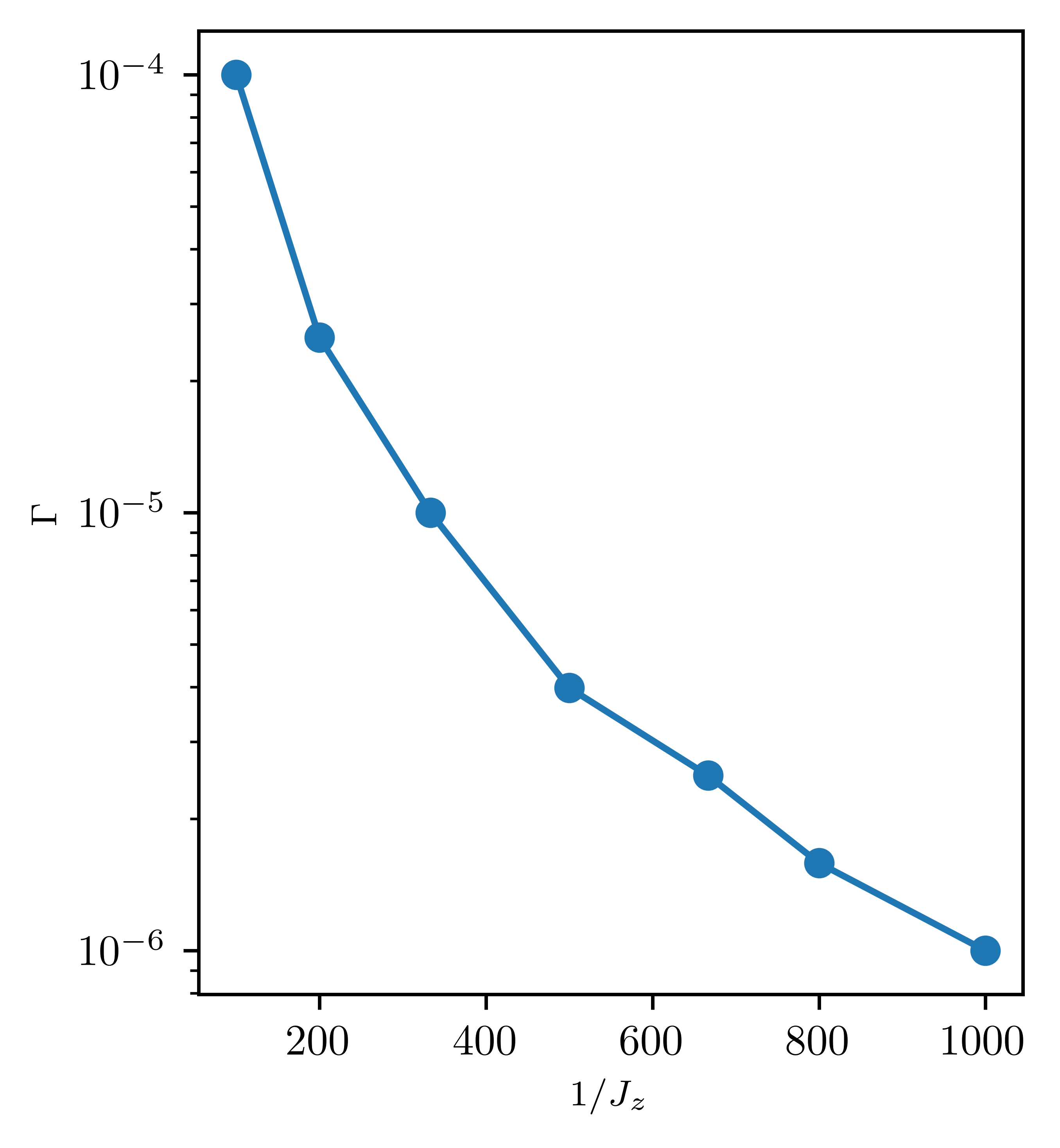}
\caption{\label{fig3a}{\bf Decay rates of the $\pi$ mode extracted from exact diagonalization (ED)}.
  The decay rate from the ED data in Fig.~\ref{fig2a} plotted on a logarithmic scale vs $1/J_z$ where $J_z$ is the interaction strength.
The constant slope for $1000\geq 1/J_z \geq 500$ indicates that
  for $1/J_z\gg 1$, the decay rate is $\Gamma \approx e^{-c/J_z}$, where $J_z$ is in units of $J_x$ and $c$
  is a $O(1)$ number.
}
\end{figure}

We employ the ansatz that $A_{\infty}^p(n T) = e^{-\Gamma  n T} A_{\infty}^p(n_0T)$, where we choose $n_0=10$ as it takes
  about 10 drive cycles for the initial transients to decay. The decay rate
  $\Gamma$ is extracted from determining the time at which $A_{\infty}^p(\Gamma^{-1}) \approx  e^{-1}A_{\infty}^p(10 T)$. This ansatz is
  plotted in Fig.~\ref{fig2a} and captures the time-evolution very well.
The decay rates $\Gamma$ are plotted in Fig.~\ref{fig3a} where now it is the $y$-axis that
is plotted on a logarithmic scale. The almost linear slope for $1/J_z \gg 1$
suggests that (restoring $J_x$)
\begin{align}
  \Gamma \sim e^{- c J_x/J_z}.\label{eqlt}
\end{align}
Above $c$ is a $O(1)$ number that depends on $g,J_x$.
Thus for small enough $J_z$, ED indicates that the decay rates are non-perturbative in the strength of the interactions $J_z$.
As $J_z$ is further increased, we do not expect the edge mode to be an ASPM, and its decay rate will be determined entirely by perturbative
processes $\Gamma \propto O(J_z^{\alpha})$, where $\alpha$ is a power of $O(1)$. 

{\bf Krylov subspace dynamics:} In order to understand the origin of the non-perturbatively long lifetimes for $J_z\ll 1$,
and its relation to bulk heating, we
map the stroboscopic time-evolution of $\sigma^x_1$ to dynamics of a free particle in a Krylov subspace
employing a recursive Lanczos scheme \cite{Recbook,Altman19,Gorsky19,Sinha19,Avdoshkin19}.
This mapping to single particle physics will allow us to develop a tunneling picture for the lifetime of the ASPM.

Let us define the Floquet Hamiltonian as
$H_F = i\ln(U)/T$. The stroboscopic time-evolution after $m$ periods
of a Hermitian operator $O$ can be written in terms of $H_F$ as follows
\begin{equation}\label{eqL}
  \left[U^{\dagger}\right]^m O U^m = e^{i H_F m T}O e^{-i H_F m T}=\sum_{n = 0}^\infty \frac{(i m T)^n}{n!} \mathcal{L}^n O,
\end{equation}
where we define
$ \mathcal{L}O = [H_F,O]$.
To employ the Lanczos algorithm, we recast the operator dynamics into
vector dynamics by defining $|O) = O$.  Since we are concerned with
infinite temperature quantities, we have an unambiguous choice for an
inner product on the level of the operators,
$  (A|B) = \text{Tr} \left[ A^\dagger B \right]/2^L$.
The Lanczos algorithm iteratively finds the operator basis that
tri-diagonalizes $\mathcal{L}$.

We begin with the seed ``state'',
$|O_1)$, and let $\mathcal{L}|O_1) = b_1 |O_2)$, where $b_1 =
\sqrt{|\mathcal{L}|O_1)|^2}$.  The recursive definition for the basis
operators $|O_{n\ge 2})$ is,
$  \mathcal{L} |O_n) = b_{n} |O_{n+1}) + b_{n-1} |O_{n-1})$,
where we define $b_{n\geq 2} = \sqrt{|\mathcal{L}|O_{n})-b_{n-1}|O_{n-1})|^2}$.  It is
straightforward algebra to check that the above procedure will
yield a $\mathcal{L}$ of the form
\begin{equation}\label{Lmat}
  \mathcal{L} =
  \begin{pmatrix}
    & b_1 &     &     & \\
    b_1 &     & b_2 &     & \\
    & b_2 &     & \ddots & \\
    &     & \ddots&   &
  \end{pmatrix}.
\end{equation}
The basis spanned by $|O_n)$ lies within the Krylov subspace of
$\mathcal{L}$ and $|O_1)$. We refer to this tri-diagonal matrix as the
Krylov Hamiltonian $H_K$,
$ H_K = \sum_n b_n ( c_n^\dagger c_{n+1} + c_{n+1}^\dagger c_n )$,
and the 1d lattice it represents, as the Krylov chain.

For free systems, the operation $\mathcal{L}|O_n)$ can be efficiently
solved in the Majorana basis. If the starting operator is a
single Majorana, then the dimension of the Krylov subspace of that
operator will scale as $2L$, as free system dynamics can only mix the
individual Majoranas among themselves. Outside of free problems,
the size of the full set of $|O_n)$ will be
large. For example, a system of size \(L\) will have $\sim 2^{2L}$
possible basis operators. Since we are interested in the thermodynamic limit for the lifetime of the edge operator, in what follows, we will treat the Krylov chain to be a semi-infinite chain.
The Krylov chain of interest to us is the one where the seed operator
$|O_1) = |\sigma_1^x)$. Then
$A_{\infty}$ is equivalent to
$  A_{\infty}(n T) = (e^{i \mathcal{L} n T})_{1,1}$.
Thus the dynamics of $A_{\infty}$ has
been transformed into that of a semi-infinite, single-particle problem where
$A_{\infty}(n T)$ is now the probability that a particle initially localized at the end of the Krylov chain,
stays localized at the end at time $n T$.

\begin{figure}[ht]
\includegraphics[width = 0.49\textwidth]{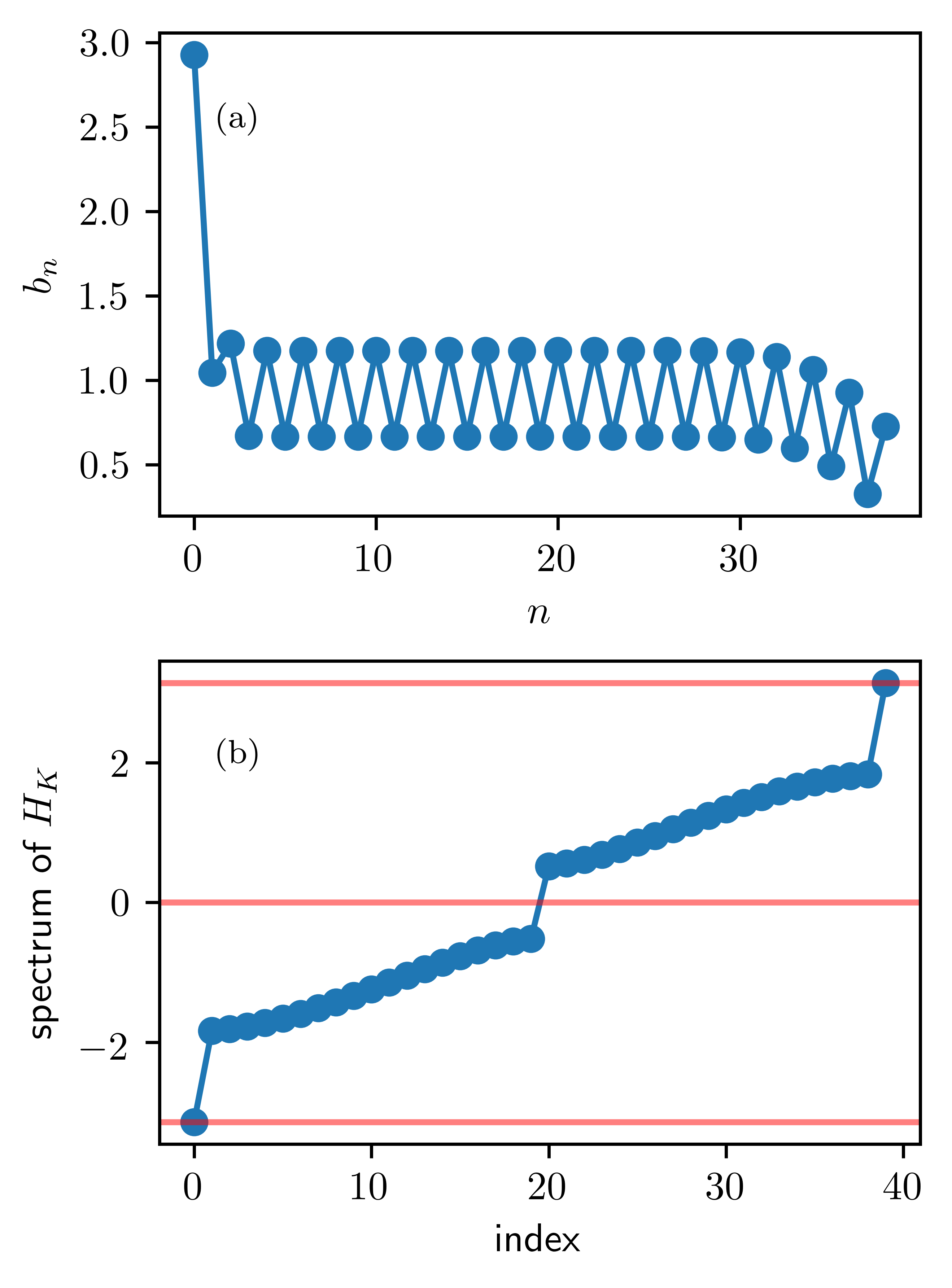}
\caption{\label{fig3a1} {\bf Parameters of the Krylov chain for the free case}. The hopping parameters $b_n$ for the Krylov chain (panel (a)) and
  the corresponding spectrum (panel (b)) for the free case $J_z=0$, where $J_z$ is the interaction strength.
  The strength of the magnetic field is $g=0.3$, the period $T=8.25$, and the system size $L=20$. $n$ labels the sites of the Krylov chain.
  The horizontal red lines in panel (b) indicate energies at $0,\pm \pi$. Panel (b) shows
  modes at $\pm \pi$ that are clearly separated from bulk states by an energy gap.
}
\end{figure}

As a point of orientation, let us discuss the details of the Krylov subspace in the free limit. In the Majorana basis, the
stroboscopic time evolution of an operator $\vec{a} =  \left[a_1, a_2, \ldots a_{2 L}\right]^T$ is (see Supplementary Note 1)
\begin{align}\label{eqfree}
  U^{\dagger}\vec{a} U = K \vec{a},
\end{align}
where $K$ is a $2L \times 2L$ orthogonal matrix and the $a_i$ are Majoranas with $a_1=\sigma^x_1$.
For studying SM dynamics, our
seed operator is $\vec{a}=\left[a_1,0,0,0\ldots\right]^T$.
The components of $K$ can be determined analytically. On
comparing equations \eqref{eqL} and \eqref{eqfree}, we identify
the operator $iT\mathcal{L}$ with $\ln{K}$. Since
${\cal L}$ is an operator, whose precise form depends on the basis, we have argued that
the Krylov Hamiltonian $H_K$ is related to $i\ln{K}$ by a simple basis rotation.

The form
of $i\ln{K}$ becomes particularly simple close to the exactly solvable point $g T=\pi$ and
in the high frequency limit $T\ll 1$. 
Denoting $s_1 = \sin(g T)$, in the first order in $s_1$ and $T$ we find (see Supplementary Note 1)
\begin{align}
  &i\ln{K} \approx \nonumber\\
  & \begin{pmatrix}
    0& i s_1& 0& 0 &0 &0 \\
    -i s_1 & 0& -i T& 0 &0& 0\\
    0&i T&0 &i s_1 & 0&0\\
    0& 0& -is_1 & 0 &-iT &0 \\
    0&0 & 0 &iT & 0 & is_1\\
    0&0 &0 & 0& -i s_1 & 0
  \end{pmatrix} \pm \pi . \label{eqMpm}
\end{align}
The analytic result in Eq.~\eqref{eqMpm} shows that $i\ln{K}$
is like a Su-Schrieffer-Heeger (SSH) model \cite{SSH79,SSH80} with a topologically non-trivial dimerization for $|s_1|< T$. Moreover the
overall shift of $\pi$ ensures that the edge mode of the SSH model is pinned at $\pi$ rather than at zero energy.
The SSH model is a band insulator with a band-gap controlled by the strength of the dimerization $||s_1|-T|$.
In contrast, when the dimerization is zero, the model is a trivial metal. We will see below
that switching on interactions leads to inhomogeneities such that insulating regions of non-zero dimerization coexist with metallic
regions of zero dimerization.

In the limit where Eq.~\eqref{eqMpm} is valid, we can derive the Krylov Hamiltonian analytically (see Supplementary Note 1). We find
  that $b_{\rm odd}= |s_1|$,
  $b_{\rm even}= T$, with a constant term $\pm \pi$ along the diagonals. Thus when $|s_1|<T$, the Krylov Hamiltonian is a topologically non-trivial
  SSH model that hosts a zero mode. The constant term along the diagonal shifts its energy to $\pi$.

The $b_n$s for $g=0.3$ and $T=8.25$, and for the free case $J_z=0$ are
shown in Fig.~\ref{fig3a1}(a). For this case, $|s_1|$ and $T$ are no longer small. Thus there are differences
in the Krylov parameters between this case, and
the exact solution around $g T \approx \pi$ just discussed. One is that the Krylov Hamiltonian has zeros on the diagonals away from the exactly solvable limit.
The second is that the hopping
on the very first site is large.  However, 
as suggested by the analytic form in the exactly solvable limit,
the Krylov chain is a SSH model with a uniform dimerization after $n\gtrapprox 4$.
Fig.~\ref{fig3a1}(b)
shows the corresponding spectrum of the Krylov chain, where the three horizontal red lines are at $\epsilon=0,\pm \pi$. Modes at $\pm \pi$ that
are also separated from the bulk modes by a gap, are clearly visible.
Thus we see that even though the diagonal term of the Krylov chain is zero, it is the 
initial large hopping of $b_1\approx \pi$ that ensures
that the edge modes of the SSH model are pinned at
$\pm \pi$.

In fact the effective model for the Krylov chain for $J_z=0$ can be written
as $H_K = H_{\rm SSH} + H_{E}$, where $H_{\rm SSH}$ represents a SSH model and captures the behavior from
sites $n\gtrapprox 4$, while $H_{E}$ is an edge Hamiltonian that captures the physics on the first few sites. The
SPM $\psi_{\pi}$ is a zero mode of $H_{\rm SSH}$, while $H_{E} \psi_{\pi} = \pi \psi_{\pi}$. Thus $H_K \psi_{\pi}= \pi \psi_{\pi}$.
To obtain a $\pi$ edge mode, the parameters of $H_E$ are finely tuned, while $H_{\rm SSH}$ only requires its dimerization to
be topologically non-trivial to ensure a zero mode. The role of $H_E$ is to raise the energy of the zero mode to $\pi$.

Note that this mapping from the Floquet unitary $U$ to the Krylov chain has lost information about the periodic nature of the spectrum
of $U$, and this manifests as finely tuned $b_n$ at the edge of the Krylov chain when a $\pi$ mode exists. Nevertheless this mapping
to an effectively free model helps to arrive at a tunneling estimate for the lifetime of the $\pi$ mode when interactions are non-zero.
We discuss this below.

\begin{figure*}
\includegraphics[width = 0.99\textwidth]{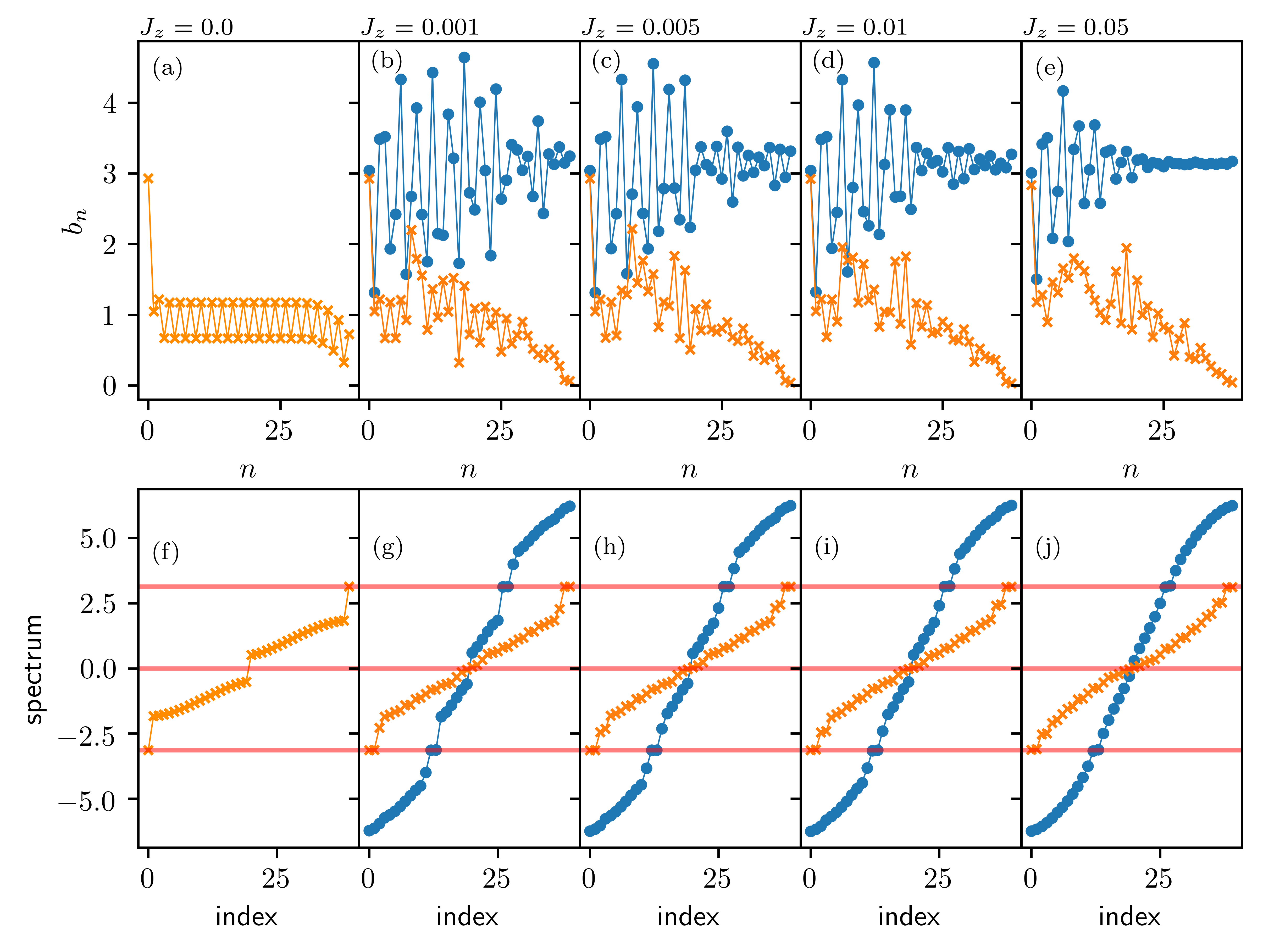}
\caption{\label{fig4a1} {\bf Interaction dependence of the parameters of the Krylov chain}.
  The Krylov chain hopping parameters $b_n$ (panels (a-e)) and the corresponding spectra (panels (f-j)), where $n$ labels the sites of the Krylov chain. 
  The results are for a magnetic field of strength $g=0.3$, period $T=8.25$, and for the interaction $J_z$ increasing from panels (a-e) and panels (f-j),
  with the $J_z=0$ case shown in panels (a,f). The system size
  for $J_z=0$ is $L=20$ (same as Fig.~\ref{fig3a1}), while that for the interacting cases is $L=12$.
  Blue circles are for the unfolded spectra, while orange crosses for non-zero $J_z$ are for a gauge choice where the
  spectra have been folded back into the Floquet Brillouin zone (FBZ). The folded spectra (orange crosses (g-j)),
  and the corresponding $b_n$ (orange crosses (b-e)) bear a closer resemblance to the
  free problem (a,f).  The horizontal red lines in panels (f-j) indicate energies at $0,\pm \pi$.
}
\end{figure*}

We now switch on interactions. We expect the dynamics to explore larger regions of the Hilbert space,
resulting in more complicated $b_n$. These are shown in Fig.~\ref{fig4a1}(b-e) for system size $L=12$
and with different $J_z$. Fig.~\ref{fig4a1}(g-j) show the corresponding spectra. For easy comparison between the free and interacting cases,
Fig.~\ref{fig4a1}(a,f) correspond to $J_z=0$.
The blue circles (all panels Fig.~\ref{fig4a1}) correspond to carrying out the Lanczos procedure in the spin basis, which is the  natural choice when
interactions are present. In contrast, the free case involved performing Lanczos in the Majorana basis Fig.~\ref{fig4a1}(a,f).
The periodicity of $U$ is lost in the Lanczos approach, and the resulting $b_n$ are sensitive to the choice of the
branch of $\ln(U)$.
This leads to $b_n$s (blue circles Fig.~\ref{fig4a1}(b-e)), which
do not bear much of a resemblance to the $b_n$s of the free case Fig.~\ref{fig4a1}(a), making them
harder to interpret. In particular, the free $b_n$s have a perfectly dimerized form for $n>3$, and therefore
a periodicity of $2$. In contrast, the $b_n$s shown by the blue circles (Fig.~\ref{fig4a1}(b-e)) have a longer periodicity, close to $3$.
The spectra for the spin basis are
shown in Fig.~\ref{fig4a1}(g-j) (blue circles). These spectra are not restricted to the FBZ.
In addition, the periodicity of $3$ manifests as $3$ gaps for the spectra shown by the blue circles (Fig.~\ref{fig4a1}(g-j)),
with the gaps located at $\pm \pi, 0$. These gaps are most
clearly visible for the smallest $J_z=0.001$ (Fig.~\ref{fig4a1}(g)).
In contrast, the spectra of the free $b_n$s (Fig.~\ref{fig4a1}(f)) have only two gaps.
The additional gap in the spectra shown by the blue circles arises because the system has lost information that the
quasi-energy spectra are continuous with $\pi$ being the same as $-\pi$.

One may map the dynamics to
an alternate Krylov subspace using an Arnoldi iteration scheme \cite{Arnoldi51}
that works directly with the Floquet unitary, rather than its logarithm, and therefore bypasses some of the ambiguities of
the Lanczos iteration. Alternatively, below we devise a scheme that can extract the relevant physics from Lanczos by a
suitable gauge choice.

Since the spectra are periodic, a physically more suitable gauge choice for the Krylov Hamiltonian is
the one where the spectra are folded back to lie within the FBZ
(orange crosses in Fig.~\ref{fig4a1}(g-j)). This folding requires
transforming the Krylov Hamiltonian $H_K \rightarrow U_K \hat{\epsilon}_{\rm FBZ} U_K^{\dagger}$, where $\hat{\epsilon}_{\rm FBZ}$
is a diagonal matrix where all the energies lie in the FBZ, and $U_K$ is the unitary matrix
that diagonalizes $H_K$ before the folding. The folding procedure is non-local, and therefore gives a new Hamiltonian, which is no longer
tri-diagonal. Therefore, a second Lanczos iteration is carried out to recover the tri-diagonal form, resulting in a new set of
$b_n$s that are shown by orange crosses in Fig.~\ref{fig4a1}(b-e).
After this transformation, the new $b_n$s bear a closer resemblance to the $b_n$s of the free case, thus making
them easier to interpret.

In comparison to the free case, one notices that a dimerization
persists even with interactions, but is non-uniform, and gradually decreases into the bulk of the chain. This is visible
in both gauges, i.e., blue circles and orange stars in Fig.~\ref{fig4a1}(a-e).
The larger $J_z$ is, the more rapidly the dimerization decays into the chain. The contrast is most visible between
$J_z=0.001$ (Fig.~\ref{fig4a1}(b)) and $J_z=0.05$ (Fig.~\ref{fig4a1}(e)). The region of the chain
where there is no dimerization, represents a metallic state. Thus we have a spatially inhomogeneous system in
the presence of $J_z$ where a disordered insulating region (represented by spatially fluctuating but non-zero dimerization)
is separated from a metallic bulk. An operator with weight in the  metallic bulk will spread rapidly, and its
autocorrelation function will decay to zero within a few drive cycles. The existence of the metal is the signature
of bulk heating because the metal has no localized states. The emergence of the metal is especially clear
after the folding procedure where the gaps at zero quasi-energy begin to fill up after the folding, compare folded spectra represented by
orange stars with the
unfolded spectra represented by blue circles in Fig.~\ref{fig4a1}(g-j).
Recall that for the free case the dimerization exists throughout the bulk.
Thus the structure of the $b_n$s for $J_z\neq 0$ gives us evidence that a quasi-localized edge mode can exist despite bulk heating.

The above picture also clarifies how the ASPM acquires a lifetime. Essentially the edge mode is localized initially
at the left end of the chain, and is separated by a finite region of dimerization from the metallic bulk.
Therefore, it has a non-zero probability
of tunneling into the metallic region. Below we estimate the lifetime of the ASPM by determining this tunneling
probability.

\begin{figure}[ht]
\includegraphics[width = 0.49\textwidth]{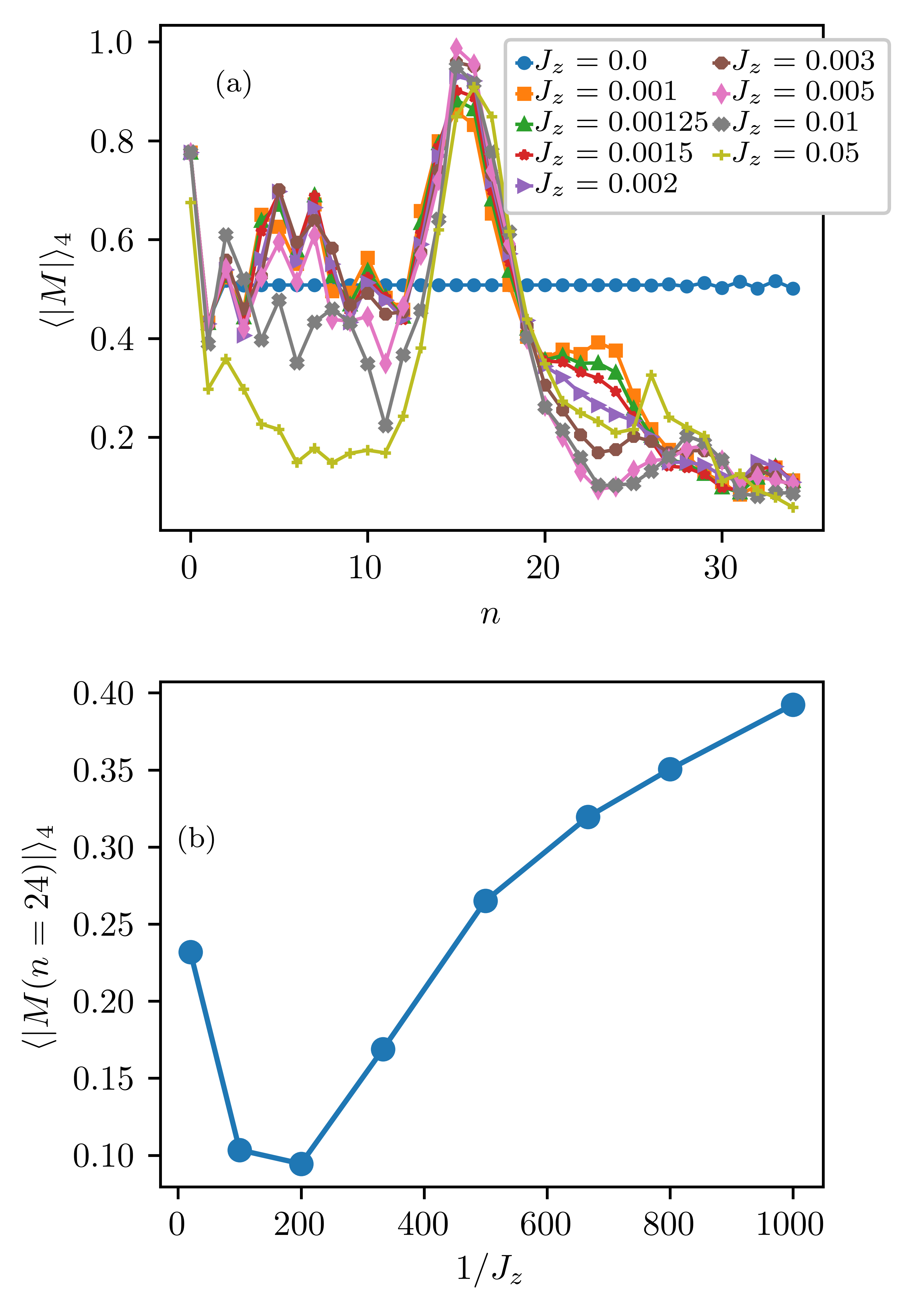}
\caption{\label{fig4a2} {\bf Interaction dependence of the dimerization of the Krylov chain}.
  The plot of the absolute value of the difference between neighboring Krylov hopping parameters
  $b_n$s, $M(n) = b_{n+1}-b_n$, after a moving average over four sites, and denoted as $\langle |M| \rangle_4$, where $n$ labels
  the sites of the Krylov chain.
  Panel (a) shows how this quantity varies with $n$ and interaction $J_z$. Panel (b) shows the variation with $1/J_z$ for
  this quantity at $n=24$. This site is chosen because the interaction effects are the strongest here. Panel (b) indicates that the
  dimerization decays with interactions as $\langle |M(n=24)|\rangle_4 \propto 1/J_z$ for $1/J_z\gg 1$.}
\end{figure}

In order to make the discussion more quantitative, in Fig.~\ref{fig4a2}(a) we plot the dimerization, i.e.,
absolute value of the nearest-neighbor $b_n$s, $M(n) = b_{n+1}-b_n$, corresponding to the data represented by the
orange stars of Fig.~\ref{fig4a1}.
We plot this
quantity after performing a moving average over 4 sites, and denote it as  $\langle |M(n)|\rangle_4$. (See Supplementary Note 2
and Supplementary Figure 1 
for the data without the averaging, and with
only 2-site averaging for comparison.)
We note that $\langle |M(n)|\rangle_4$ does not change
with $J_z$ for the first couple of sites (provided $J_z<0.05$),
while away from the edge, $\langle|M(n)|\rangle_4 $
decreases with $n$ when $J_z\neq 0$. In contrast, $\langle|M(n)|\rangle_4 $ stays constant for the free case.
The fact that the first few sites of the Krylov chain do not change with $J_z$ implies that $H_E$ is not sensitive to
$J_z$. We therefore adopt a simple model of a Krylov chain for the sites $n\gtrapprox 4$,
with two slowly varying parameters, a
nearest-neighbor average hopping $(b_n + b_{n-1})/2$, and the dimerization $M(n)$ \cite{Yates20, Yates20a} (see Supplementary Note 3).

We now emphasize some important differences between the $b_n$s for the ASZM in static systems
\cite{Yates20, Yates20a} and the same for the ASPM for Floquet systems. One is the sensitivity to gauge choice for the latter due to
the freedom in shifting the
quasi-energies by integer multiples of $2\pi$ (see detailed discussion above).  The
second difference is that when the Floquet spectrum is bounded by the FBZ, the average $b_n$s do not increase unboundedly with $n$,
unlike in static systems.
Thus we derive a continuum model under the assumption that the nearest-neighbor average hopping is spatially uniform, and that the
dimerization is slowly varying in space. These assumptions map the Krylov chain onto
a Dirac model with a spatially inhomogeneous mass (see Supplementary Note 3)
$i\partial_t \tilde{\Psi} = \left[m(X) \sigma^y + \sigma^x i\partial_X\right]\tilde{\Psi}$,
where $m(n)= M(2n)$. For $J_z=0$, the mass is spatially uniform and topologically non-trivial, $m>0$, with
$M(2n)= -M (2n+1) = m$.
For $J_z\neq 0$, this mass vanishes into the bulk. While the precise model
for how it vanishes is complicated, we adopt a simple ansatz where $m(X) = M_0 \theta(X-X_0)$. Then a WKB treatment shows that the
lifetime of the edge mode is (see Supplementary Note 3) $\Gamma \approx 4 M_0 e^{-2 M_0 X_0}$.
The fact that the edge mode is at $\pi$ energy rather than at zero energy enters in the boundary condition via
$H_E$, where a strong local hopping pins the edge mode to $\pi$.

We now discuss the $J_z$ dependence of $M_0$.
Fig.~\ref{fig4a2}(b) shows that $\langle |M(n=24)|\rangle_{4} \sim J_z^{-1}$. Since the
decay-rate $\Gamma$ depends on the mass $M_0$ exponentially, and $M_0 \propto 1/J_z$, we conclude that
$\Gamma \sim e^{-c/J_z}$.

{\bf Bound for lifetime from domain wall counting argument:}
We now present an alternate argument for the non-perturbatively long lifetime in Eq.~\eqref{eqlt}.
We show below that despite the low frequency driving, the energy required to
flip the spin on the very first site is highly off-resonant, and requires
rearranging many domain walls in the bulk. This phenomena was also noted
in previous studies on static systems \cite{Else17,Nayak17,Laumann20,Yates20}.
Thus the boundary is in a prethermal state
\cite{Abanin15,Kuwahara2016,Mori16,Abanin17,Abanin17b}, despite the thermalized
bulk.
We argue this physics by deriving a Floquet Hamiltonian $H_F$ in the limit of
$J_z\ll 1$ and $|g T-\pi|\ll 1$. Recall $gT = \pi$ and $J_z=0$ is the exactly solvable limit where the SPM
is $\Psi_\pi = \sigma^x_1$.

Let us define \( \hat{J}_z =J_zT/2, \hat{g}= T g/2- \pi/2\) and $ \hat{J}_x =T J_x/2 $. We will
work in the limit $\hat{J}_x\gg 1$ and $\hat{J}_z,\hat{g} \ll 1$.
We cannot perform a high-frequency expansion \cite{Eckardt15} to construct $H_F$ as \(\hat{J}_x\) is not
small. Nevertheless, \(H_F\) to first order in
\(\hat{J}_z,\hat{g}\) but to arbitrary orders in \(\hat{J}_x\) may be
derived from an infinite resummation of the Baker-Campbell-Hausdorff
formula \cite{Scharf1988,Polkov13}, leading to the following non-local perturbed Ising model \cite{Yates19}
\begin{widetext}
\begin{align} \label{HFBC}
  TH_F &\approx \hat{J}_x H_{xx} + \frac{\pi}{2} \mathcal{D} + \hat{g} \biggl\{  h_{z}^E \hat{J}_x \cot(\hat{J}_x)
          + h_{z}^B\left( \frac{1 + 2 \hat{J}_x \cot(2\hat{J}_x)}{2} \right)
          + h_{xzx}\left( \frac{-1+ 2 \hat{J}_x \cot(2\hat{J}_x)}{2} \right)
          -\hat{J}_x \left( h_{xy} + h_{yx} \right)\biggr\}\nonumber\\
    &+ \hat{J}_z \biggl\{  h_{zz}^E \hat{J}_x \cot(\hat{J}_x)
           + h_{zz}^B\left( \frac{1 + 2 \hat{J}_x \cot(2\hat{J}_x)}{2} \right)
          - h_{xyyx}\left( \frac{-1 + 2 \hat{J}_x \cot(2\hat{J}_x)}{2} \right)
           + \hat{J}_x \left( h_{zyx} + h_{xyz} \right)\biggr\} + O(\hat{g}^2,\hat{J}_z^2).
\end{align}
\end{widetext}
Above
\(h_{\alpha_1\dots\alpha_k}=\sum_j\sigma_j^{\alpha_1}\dots\sigma_{j+k-1}^{\alpha_k}
\equiv h^E_{\alpha_1\dots\alpha_k}+h^B_{\alpha_1\dots\alpha_k}\), where $h^E$
denotes the contributions from the edge spins $\sigma_{1,L}^\alpha$ and $h^B$ denotes the bulk spins.

Let us consider the energetics involved in flipping $\sigma^x_1$ for $\hat{g}=\hat{J}_z=0$. Since the
boundary spin has only one neighbor, this flip costs energy $\hat{J}_x$.
However, the energy cost for creating a domain wall in the bulk is $2\hat{J}_x$ due to the two neighboring sites. Thus there
is a mismatch of $\hat{J}_x\gg 1$ between a bulk and an edge excitation, making the
flipping of an edge spin impossible. However this simple argument does not account for
processes that can make the domain wall hop from site to site, resulting in a lowering of
its energy. Thus, we have to revisit the energy argument by accounting for the kinetic
energy of the domain walls.

In order to develop our argument, we first note
that $H_{xx}$ counts the number
of domain-walls $N=\sum_{i}\sigma^x_i \sigma^x_{i+1}$. Therefore, we write $H_F$ as  a part
that commutes with the number of domain walls, and a part that
changes the number of domain walls. In particular,
$T H_F=\hat{J}_x N + D +V$ where $\left[D,N\right]=0$ and $\left[V,N\right]\neq 0$.
Both $D$ and $V$ can be written in the form similar to Eq.~\eqref{HFBC}, i.e., as sums over local strings of
Pauli matrices. The precise forms of $D$ and $V$ are not needed for our qualitative argument.
We only assume that the operator norms of each of the local terms of $D$ and $V$ are $O(\hat{g},\hat{J}_z)$.
If these assumptions are valid
we can apply the arguments of Refs.~\cite{Else17,Nayak17,Laumann20,Yates20} to find a lower bound on the lifetime of the ASPM.
Here, we briefly summarize the physical picture.

First, consider the Hamiltonian $\hat{J}_x N + D$. The spectrum of the $\hat{J}_x$-term are states that
are separated by multiples of $\hat{J}_x$ because $N$ counts domain walls.
The $D$-term causes the domain walls to move without changing their number.
Diagonalization of $\hat{J}_x N + D$ results in domain wall ``bands'' with a typical bandwidth
$\epsilon\sim ||D||\sim O(\hat{g},\hat{J}_z)$ which is much smaller than the separation $\hat{J}_x$ between the bands.

The $V$-term of the total Hamiltonian does change the number of domain walls, but only by an even number due to the parity symmetry
of the total Hamiltonian. A single application of $V$ therefore changes the energy by about $2\hat{J}_x$ and is off resonant
with the cost $\hat{J}_x$ of flipping the boundary spin.
It is impossible to absorb the energy $\hat{J}_x$ within few orders of perturbation theory in $V$. However, the creation and annihilation
of a pair of domain walls would lead to the change of the energy by the order of the bandwidth
$\epsilon\ll \hat{J}_x$.  Therefore, we estimate that one needs
of the order of  $\hat{J}_x/\epsilon$ powers of $V$ to offset the energy $\hat{J}_x$ required to flip a boundary spin.
The probability corresponding to the required order of perturbation theory goes as
$\left[||V||\right]^{\hat{J}_x/\epsilon}\sim \left[||V||\right]^{J_x/O(J_z,g)}$
where $||V||$ denotes the typical size of the matrix element that creates a domain wall. The above expression
is a lower bound for the lifetime. For example, in the two integrable limits (which is a property of the exact rather than the
approximate $H_F$) $J_z\to 0, g\neq 0$ and $g\to 0, J_z \neq 0$, the lifetime
should diverge.
Empirically combining this observation with the rough estimate above we expect $1/\epsilon = O(1/J_z, 1/g)$.
When $J_z \ll g$ this empirical formula replaces $\epsilon \to J_z$ in the above estimate making it consistent with Eq.~\eqref{eqlt}.

{\bf Conclusions:}
ASPMs are fascinating objects which have lifetimes that far exceed bulk heating times. Besides presenting
evidence for this, we developed a method for extracting their lifetimes by mapping their
dynamics to single-particle quantum mechanics in Krylov subspace. While we studied the lifetime for $J_z\ll g$,
determining the lifetime when $g,J_z$ are comparable is left for future studies.
Our Krylov method for determining lifetimes is generalizable to any spatial dimension, to closed and open
systems, and to static and driven systems.
In addition, the resistance to heating of the $\pi$ mode is promising for its experimental realization \cite{Liu19}.\newline
{\bf Data Availability:} All relevant data are available from the corresponding author upon reasonable request.\newline
{\bf Author contributions:} DJY and AM performed the numerical and analytical calculations. AGA and AM helped in the interpretation of the results and
in the writing of the manuscript.  \newline
{\bf Competing Interests:} The authors declare no competing interests.\newline 
{\bf Acknowledgements:}
This work was supported by the US Department of Energy, Office of
Science, Basic Energy Sciences, under Award No.~DE-SC0010821 (DJY and AM)
and by the US National Science Foundation Grant NSF
DMR-1606591 (AGA).


%

\begin{widetext}
\setcounter{figure}{0}
\setcounter{equation}{0}
\setcounter{page}{0}
\renewcommand{\thepage}{S\arabic{page}} 
\renewcommand{\thesection}{S\arabic{section}}  
\renewcommand{\thetable}{S\arabic{table}}  
\renewcommand{\thefigure}{S\arabic{figure}}
\renewcommand{\theequation}{S\arabic{equation}}

\section*{Supplementary Note 1: Derivation of Eq.~(8) and the corresponding Krylov hopping parameters}

For the binary drive corresponding to setting $J_z=0$ in Eq.~(1) in the main text, it is convenient to map the problem
to Majorana fermions as follows
\begin{align}
  a_{2\ell -1} = \prod_{j=1}^{\ell-1}\sigma^z_j \sigma^x_\ell;\,\, a_{2\ell} = \prod_{j=1}^{\ell-1}\sigma^z_j \sigma^y_\ell,
\end{align}
where $\ell = 1 \ldots L$.
Denoting the vector
\begin{align}
  \vec{a} = \begin{pmatrix} a_1 \\ a_2 \\ a_3 \\\vdots \\a_{2L}\end{pmatrix},
\end{align}
and for a system of size $L$, the time-evolution of
an operator $\vec{a} = \left[a_1, a_2,a_3,\ldots a_{2L}\right]^T$ is given by Eq.~(7) in the main text. The explicit form for $K$ for
a system of size $L=3$ is \cite{Yates19}
\begin{align}
   K&=
  \begin{pmatrix}
    c_1& -s_1& 0& 0 &0 &0 \\
    s_1c_2 & c_1 c_2& -c_1s_2& s_1 s_2 &0& 0\\
    s_1 s_2&c_1s_2&c_1c_2 &-s_1c_2 & 0&0\\
    0& 0& s_1 c_2 & c_1c_2 &-c_1 s_2 & s_1 s_2 \\
    0&0 & s_1 s_2 &c_1s_2 & c_1c_2 & -s_1c_2 \\
    0&0 &0 & 0&s_1 & c_1
  \end{pmatrix}.
  \label{eqM}
\end{align}
where,
\begin{subequations}\label{eqcsdef}
  \begin{align}
    c_1 &= \cos(Tg),\\
    c_2 &= \cos(T),\\
    s_1 &= \sin(Tg),\\
    s_2 &= \sin(T).
  \end{align}
\end{subequations}
$K$ is a real orthogonal matrix as it represents rotations in the Majorana basis,
\begin{align}
  K^T K = 1.
\end{align}
$K$ can be easily generalized to larger system sizes by noting that the second and third rows keep repeating (upto a shift by two columns
to the right)
until the last row i.e, the $2L$-th row. 

An exactly solvable point for the $\pi$ modes is $T g=\pi$ \cite{Khemani16}.
Denoting $K=K_0$ when $T g =\pi$, and noting that at this point
$s_1=0,c_1=-1$, we have
\begin{align}
& K_0=
  \begin{pmatrix}
    -1& 0& 0& 0 &0 &0 \\
    0 & -c_2& s_2& 0 &0& 0\\
    0& -s_2& -c_2 &0 & 0&0\\
    0& 0& 0 & -c_2 &s_2 & 0 \\
    0&0 & 0 & -s_2 & -c_2 & 0 \\
    0&0 &0 & 0&0 & -1
   \end{pmatrix}\nonumber\\
  & =
\left(
    \begin{tabular}{ c c c c}
 -1 & \begin{tabular}{c c}
   \ \ 0 &  \! \! \ \ 0 \
 \end{tabular} & \begin{tabular}{c c}
     0 & \! \! \ \ 0
 \end{tabular} & \ 0 \\
 \begin{tabular}{c}
     0 \\
     0
 \end{tabular} & \(-e^{-iT\sigma_y}\) & \begin{tabular}{c c}
     0 & \! \! \ \ 0   \\
     0 & \! \! \ \ 0
 \end{tabular} & \begin{tabular}{c}
     \ 0 \\
     \ 0
 \end{tabular} \\
 \begin{tabular}{c}
     0 \\
     0
 \end{tabular} & \begin{tabular}{c c}
      \ \ 0 &  \! \! \ \ 0 \  \\
      \ \ 0 &  \! \! \ \ 0 \
 \end{tabular} & \(-e^{-iT\sigma_y}\) & \begin{tabular}{c}
     \ 0 \\
     \ 0
 \end{tabular} \\  0 & \begin{tabular}{c c}
      \ \ 0 &  \! \! \ \ 0 \
 \end{tabular} & \begin{tabular}{c c}
     0 & \! \! \ \ 0
 \end{tabular} & \ -1
\end{tabular}\right).
  \label{Mgpi}
\end{align}
For later use, we also write the inverse of $K_0$ explicitly,
\begin{align}
  K_0^{-1} =
  \begin{pmatrix}
    -1& 0& 0& 0 &0 &0 \\
    0 & -c_2& -s_2& 0 &0& 0\\
    0& s_2& -c_2 &0 & 0&0\\
    0& 0& 0 & -c_2 &-s_2 & 0 \\
    0&0 & 0 & s_2 & -c_2 & 0 \\
    0&0 &0 & 0&0 & -1
   \end{pmatrix}.
\end{align}

Taking the logarithm of the above quantity, one obtains
\begin{align}
\ln{K_0} &= -i T \begin{pmatrix}
    0& 0& 0& 0 &0 &0 \\
    0 & 0& -i& 0 &0& 0\\
    0&i&0 &0 & 0&0\\
    0& 0& 0 & 0 &-i & 0 \\
    0&0 & 0 &i & 0 & 0 \\
    0&0 &0 & 0&0 & 0
  \end{pmatrix} \pm i\pi.\label{eqM0mpm}
\end{align}
The first part of the equation hosts a zero mode, while the term $\pm \pi$ shifts the energy to $\pm \pi$, giving a strong $\pi$ mode (SPM).

We now study the effect of small deviations from the exactly solvable limit $gT=\pi$. Let us write, $K = K_0 + V_p$ where
\begin{align}
 &  V_p= 
  \begin{pmatrix}
    0& -s_1& 0& 0 &0 &0 \\
    s_1c_2 & 0& 0& s_1 s_2 &0& 0\\
    s_1 s_2&0&0 &-s_1c_2 & 0&0\\
    0& 0& s_1 c_2 & 0 &0 & s_1 s_2 \\
    0&0 & s_1 s_2 &0 & 0 & -s_1c_2 \\
    0&0 &0 & 0&s_1 & 0
  \end{pmatrix} + O(s_1^2).
\end{align}
If in addition to imposing $|s_1|\ll 1$, we also impose that $T\ll 1$ (or $|s_2|\ll 1$), then
$K_0$ and $V_p$ commute if terms of $O(s_1s_2), O(s_1^2), O(s_2^2)$ and higher are neglected.
Recall that if $A$ and $B$ are two commuting matrices $\left[A,B\right]=0$, then,
\begin{align}
  \ln{(A+B)} \approx \ln{A} + A^{-1} B + \ldots. \label{eqlogexp}
\end{align}
Neglecting terms of $O(s_1s_2), O(s_1^2), O(s_2^2)$ and higher, we have two commuting matrices $A=K_0, B=V_p$. We can then use
Eq.~\eqref{eqlogexp} 
to obtain the following expression to first order in $s_1$ and $T$
\begin{align}
  &i\ln{K} \approx
  \begin{pmatrix}
   0 & i s_1& 0& 0 &0 &0 \\
    -i s_1 & 0& -i T& 0 &0& 0\\
    0&i T&0 &i s_1 & 0&0\\
    0& 0& -is_1 & 0 &-iT &0 \\
    0&0 & 0 &iT & 0 & is_1\\
    0&0 &0 & 0& -i s_1 & 0
  \end{pmatrix} \pm \pi . \label{eqMpm2}
\end{align}
The above is Eq.~(8) of the main text.

We now carry out the Krylov procedure. We will show that the Liouvillian in the Krylov basis is a tridiagonal matrix of the form
\begin{align}
  {\cal L} = \begin{pmatrix}
    & b_1 &     &     & \\
    b_1 &     & b_2 &     & \\
    & b_2 &     & \ddots & \\
    &     & \ddots&   &
  \end{pmatrix} \pm \pi.
\end{align}

Without loss of generalization, let us choose the $+$ sign in $\pm \pi$ in Eq.~\eqref{eqMpm2}, and also take $s_1>0$.
The first operator is $\sigma^x_1=a_1$, which is represented by the
vector,
\begin{align}
  |O_1) = \left[1,0,0,0,0,0\right]^T.
\end{align}
We generate the second operator, or vector as follows,
\begin{align}
  |O_2') &= i \ln{K} |O_1)= \left[\pi ,-i s_1,0,0,0,0\right]^T. 
\end{align}
The diagonal term of the Liouvillian is
\begin{align}
  \alpha_1 = (O_1|i\ln{K}|O_1)=(O_1|O_2') = \pi.
\end{align}
Now we orthogonalize the state $|O_2')$ with respect to the previous state as follows,
\begin{align}
  |O_2'') = |O_2') - \alpha_1 |O_1) = \left[0,-i s_1,0,0,0,0\right]^T.
\end{align}
The first hopping term in the Liouvillian is
\begin{align}
  b_1 = \sqrt{(O_2''|O_2'')} = s_1,
\end{align}
and normalizing the resulting state we obtain
\begin{align}
  |O_2) = |O_2'')/b_1 =\left[0,-i,0,0,0,0\right]^T.
\end{align}
Continuing this process, we generate the next state by the application of $i\ln{K}$ to the previous state,
\begin{align}
|O_3')&= i \ln{K} |O_2)\nonumber\\
  &= \left[s_1, -i\pi ,T,0,0,0\right]^T. 
\end{align}
The diagonal term of the Liouvillian is
\begin{align}
  \alpha_2 =(O_2|i\ln{K}|O_2) =(O_2|O_3') = \pi,
\end{align}
and orthogonalizing with respect to the previous two states, we obtain
\begin{align}
  |O_3'') = |O_3') - \alpha_2 |O_2) - b_1 |O_1)= \left[0,0,T,0,0,0\right]^T.
\end{align}
Thus the second hopping term is
\begin{align}
  b_2 = \sqrt{(O_3''|O_3'')} = T,
\end{align}
and the new state after normalization is 
\begin{align}
  |O_3) = |O_3'')/b_2=\left[0,0,1,0,0,0\right]^T.
\end{align}
Repeating the process, 
\begin{align}
|O_4')&= i \ln{K} |O_3)\nonumber\\
  &= \left[0, -i T ,\pi, -i s_1,0,0,0\right]^T. 
\end{align}
The diagonal term is
\begin{align}
  \alpha_3 = (O_3|O_4') = \pi,
\end{align}
and orthogonalizing with respect to $|O_{1,2,3})$ we have
\begin{align}
  |O_4'') = |O_4') - \alpha_3 |O_3) - b_2 |O_2)= \left[0,0,0,-i s_1,0,0\right]^T.
\end{align}
Thus,
\begin{align}
  b_3= \sqrt{(O_4''|O_4'')} = s_1,
\end{align}
and
\begin{align}
|O_4) = |O_4'')/b_3 = \left[0,0,0,-i,0,0\right]^T.
\end{align}

We now see a straightforward pattern emerge where $b_{\rm odd} = |s_1|$, $b_{\rm even}=T$, with diagonal terms $\alpha_i =\pi$.
A zero mode exists for the topologically non-trivial dimerization
$|s_1|<T$. Due to the $\pi$ term along the diagonals, the energy of the mode is increased to $\pi$.

\section*{Supplementary Note 2: Dimerization and its moving average}

For the free case ($J_z=0$), the absolute value of the dimerization, defined as the absolute value of the 
difference between neighboring $b_n$s, $|M(n)|=|b_{n+1}-b_n|$,
is constant after the first few sites. However when interactions are added, the dimerization
fluctuates from site to site. To obtain a physically meaningful quantity, in the main text, we performed a four site moving average of the absolute value of the
dimerization. Here, Supplementary Figure 1(a)
shows how the absolute value of the dimerization behaves without any averaging,
while Supplementary Figure 1(b) shows how this quantity behaves after a two site moving average.
Note that the main features of the dimerization, namely that it is now non-zero only over a finite spatial region of size
$n \approx 25$, beyond which it decays to zero, is preserved in the averaging procedure. 

\begin{figure}[ht]
  \includegraphics[width = 0.6\textwidth]{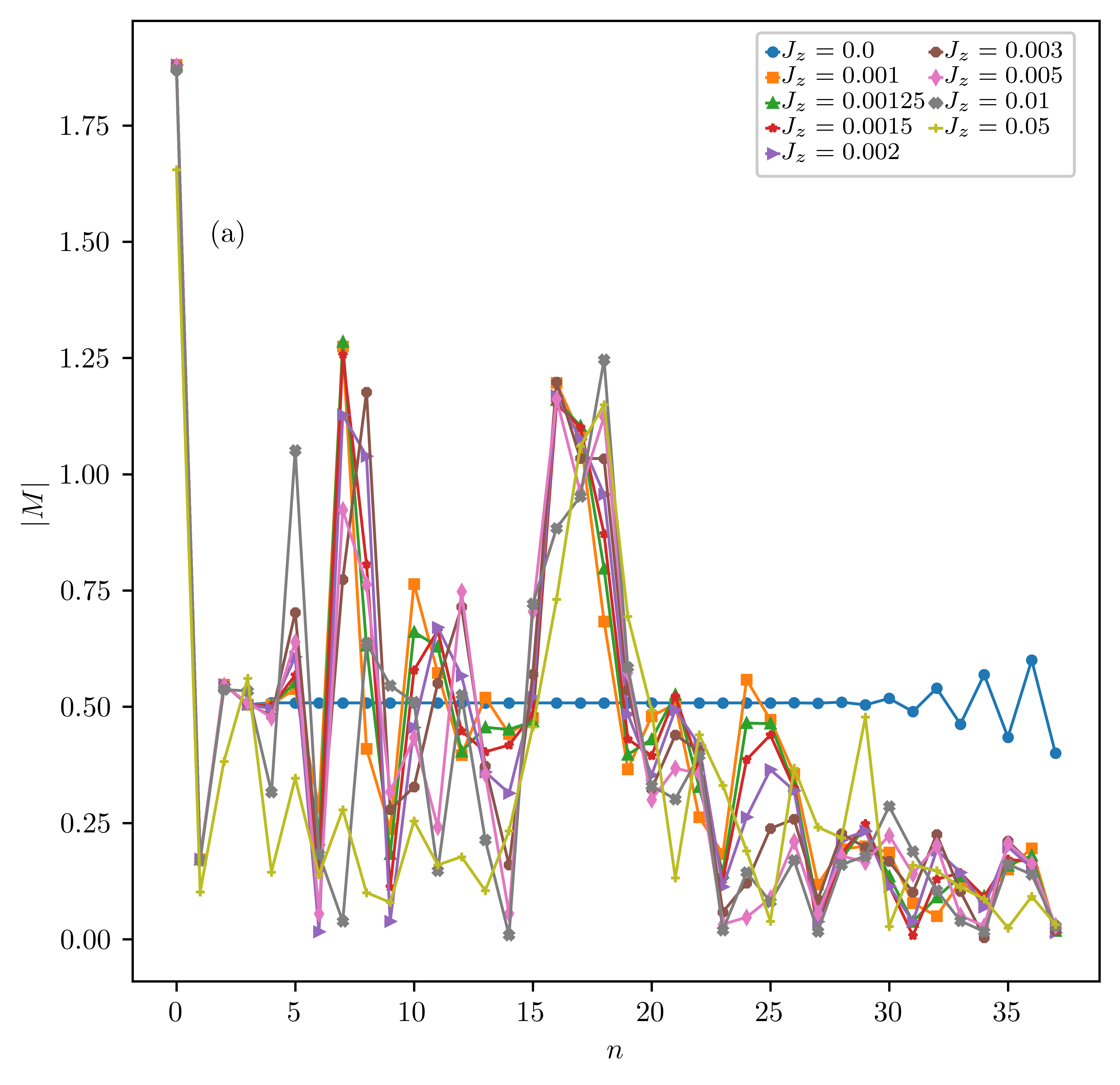}
  \includegraphics[width = 0.6\textwidth]{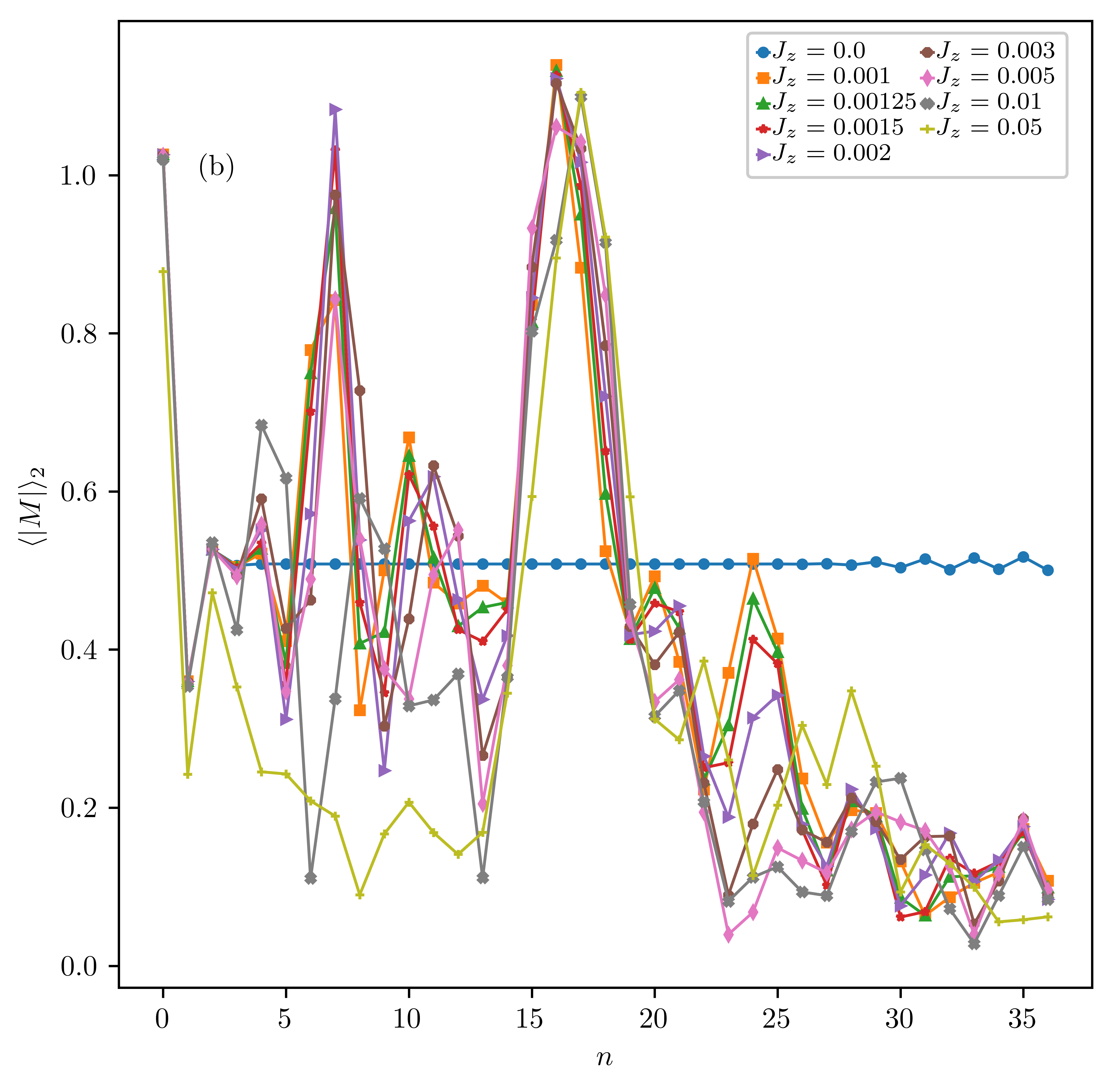}
  \caption{
    {\bf Supplementary Figure 1: Absolute value of the dimerization and its moving average.}
Panel (a): The plot of the absolute value of the difference between neighboring
$b_n$s, $|M(n)| = |b_{n+1}-b_n|$. Panel (b): The plot of the absolute value of the difference between neighboring
$b_n$s, and averaged over 2 sites. This quantity is denoted as $\langle |M| \rangle_2$. The main text shows the
same quantity, but after a 4-site averaging. One notes that when $J_z\neq 0$, $|M|$ fluctuates with position $n$. However, it is non-zero upto
about $n=25$, beyond which it decays to zero. In addition, 
as $J_z$ increases, the average dimerization decreases.
}
\end{figure}

\section*{Supplementary Note 3: Mapping to a Dirac model with inhomogeneous mass}

The Schr\"{o}dinger equation for the edge mode in Krylov subspace can be written as
\begin{equation}
  i \partial_t \Psi_n = b_n \Psi_{n+1} + b_{n-1} \Psi_{n-1}.\label{se1a}
\end{equation}
Note that we are considering a semi-infinite system where $n\geq 1$ with $b_0=0$.
Since any diagonal term is a constant of the form ${\rm integer}\times \pi$, we drop it here, as its effect would be to simply 
shift the energy of the edge mode by the constant amount. 

To transform to the continuum limit we assume that the hopping
parameters and the wavefunction can be written as
\begin{eqnarray}
  \Psi_n &=& i^n\left[\psi_n + (-1)^n \tilde{\psi}_n \right],
\label{psian} \\
  b_n &=& h_n + (-1)^n\tilde{h}_n,
 \label{bnan}
\end{eqnarray}
where $\psi_n,\tilde\psi_n,h_n,\tilde{h}_n$ are all assumed to be
smooth, slowly varying functions of $n$.
We now measure distances in lattice spacings $x=n$, introduce continuous
notations $\psi_n=\psi(x)$ etc., and introduce the spinor,
\begin{equation}
  \tilde{\Psi} =
  \begin{pmatrix}
    \psi(x) \\ \tilde{\psi}(x)
  \end{pmatrix}.
\end{equation}

In Eq.~\eqref{se1a} we substitute the ansatz Eq.~\eqref{psian} for the
wavefunction, and the ansatz Eq.~\eqref{bnan} for the hopping
amplitudes. Following the derivation outlined in Ref.~\onlinecite{Yates20a} which assumes
that $h, \tilde{h}, \psi, \tilde{\psi}$ are smoothly varying functions of $x$, we arrive at,
\begin{equation}\label{cont2}
  i \partial_t \tilde{\Psi}(x) =
  \left[ \sigma^y m(x) +
    \sigma^z\{i\partial_x,h(x)\} \right] \tilde{\Psi}(x).
\end{equation}
Above, the mass is defined as
\begin{equation}\label{m1def}
  m(x) = 2\tilde{h}(x) - \partial_x\tilde{h}(x).
\end{equation}

An important difference between the static problem \cite{Yates20a}, and the Floquet problem studied here is that
since the Floquet spectrum is bounded, the average of the nearest-neighbor hoppings, ${h}(x) \approx $ constant.
In the static case ${h}(x) \propto x$, and therefore grows unboundedly as $x$ increases.
Under the assumption of a constant ${h}(x)=h$, we arrive at,
\begin{align}
	i\partial_{t}\tilde{\Psi}(X,t) &= \biggl[ \sigma^{z}i\partial_{X}+\sigma^{y} m(X)\biggr]\tilde{\Psi}(X,t),
\end{align}
where the mass is space dependent, and we defined a rescaled coordinate $X = x/2h$.

To capture the main physics of the non-perturbative effect of $J_z$, it is sufficient to make the assumption that
$m(X)=M_0=\rm{const}>0$ for $0\leq X \leq X_{0}$ and $m(X)=0$ for $X>
X_{0}$. For this case, after a straightforwardly application of the WKB treatment, one finds the decay rate to be
\cite{Yates20a},
\begin{align}
	\Gamma \approx 4M_0\, e^{-2M_0X_{0}}\,.
 \label{Gamma100}
\end{align}

\end{widetext}
  
\end{document}